\documentstyle[12pt,epsf,cite]{article}
\newcommand{\be}{\begin{equation}}
\newcommand{\ba}{\begin{eqnarray}}
\newcommand{\ee}{\end{equation}}
\newcommand{\ea}{\end{eqnarray}}
\newcommand{\nn}{\nonumber}

\newcommand{\MeV}{\;\mbox{MeV}}
\newcommand{\GeV}{\;\mbox{GeV}}
\newcommand{\TeV}{\;\mbox{TeV}}
\newcommand{\simlt}{\stackrel{<}{{}_\sim}}

\newcommand{\ol}{\overline}

\begin{document}
\title{Unusual behavior in $\overline{p}-p$ and
${\ell}-p$ collisions involving high energy and momentum transfer:
$Z^0(W^\pm)$ and scalar jets.}
\author{Saul Barshay and Georg Kreyerhoff\thanks{E-mail: georg@physik.rwth-aachen.de}\\
III. Physikalisches Institut A\\
RWTH Aachen\\
D-52056 Aachen\\
Germany}
\maketitle
\vspace{-10cm}
\begin{flushright}PITHA 97/15\\
(addended)\end{flushright}
\vspace{10cm}
\begin{abstract}
We show that the axial-vector coupling of $Z^0(W^\pm)$ to quarks,
when acting together with the emission of a hypothetical spin-zero
jet-generating quantum coupled to quarks results in an unusual behavior
in certain processes, at high energy and momentum transfer. This involves
an excess jet production at the subpicobarn level.
\end{abstract}
The purpose of this paper is to show how a general, dynamical effect
can result in unusual behavior at high energy and momentum transfer,
in processes which involve jet production and a real, or virtual,
$Z^0(W^\pm)$. We present definite numerical examples of the effect 
which are applicable to $\ol{p}-p$ collisions\cite{ref1}, and to
${\ell}-p$ collisions\cite{ref2}, where data is being collected and analyzed.
The specific motivations for emphasizing this general, dynamical phenomenon
are the following.
\begin{itemize}
\item[(1)] There is a possible excess \cite{ref3,ref4} in the production of
$W$ plus one jet, in $\ol{p}-p$ collisions at 1.8 TeV at the
Tevatron.\cite{ref4}.
\item[(2)] There is also a possible excess in the production of $Z$ plus
one jet at Fermilab\cite{ref5}. In determining the expectation from 
QCD, the precise choice of scale seems to be relevant\cite{ref5}.
\item[(3)] There may be an excess in the production of $W$ plus two jets at
Fermilab\cite{ref5}.
\item[(4)] A small excess in inclusive jet production may be present
in the Tevatron experiments\cite{ref1,ref4,ref6,ref7}. The range of $x$-values
and of $p_T$ are rather similar to the range of these essential kinematic
variables (high initial collision energy and high final transverse energy),
relevant to the HERA anomalous events, the next point.   
\item[(5)] In positron-proton collisions at HERA there is a possible excess
\cite{ref2} in the jet production associated with deep-inelastic
scattering.
\end{itemize}
In the latter experiments, the negative of the squared 
four-momentum transfer from the lepton is $(-Q^2) \sim (2m_Z)^2$;
thus virtual $Z(W)$ exchange is comparable to virtual photon exchange.
Then, the processes (1),(2),(3) and (5) all involve explicitly a massive vector
boson with its axial-vector coupling to quarks, and a hadronic jet, or jets.
It is of interest to examine a general, hypothetical mechanism which
may be capable of giving a common physical basis for these several
effects, which await experimental confirmation.\par
Such a general mechanism arises when the axial-vector coupling of
a massive vector boson to quarks acts in a matrix element, together
with the scalar (or pseudoscalar) coupling to quarks of a hypothetical
spin-zero, jet-generating quantum $J_0$\cite{ref8,ref9}.
The essential point is illustrated by considering the process $\ol{q} + q
\to Z+J_0$. The two Feynman diagrams are shown in fig.~1. Instead
of emission of the colored gluon with spin-one of QCD, we consider
emission of an analogous, hypothetical spin-zero quantum, thus 
replacing $\gamma_\mu$ by 1 at the emission vertex and denoting the
coupling constant by $g_0$. The differential cross section is
\be
\frac{d\sigma_{J_0 Z}}{dy} = \sqrt{2}G_F \left(\frac{4}{9}\frac{g_0^2}{4\pi}\right)
\left( 1-\frac{m_Z^2}{4E^2}\right)\left[\frac{m_Z^2}{4E^2}\left(
\frac{A}{1-y^2} - B\right)+\frac{B}{2}\right]
\ee
The total energy of the $\ol{q}-q$ annihilation is $2E$ and the emission angle
of $J_0$ is $\theta$, with $y=\cos\theta$, in the center-of-mass (CM) system of
the $\ol{q}-q$ collision. In eq.~(1), we have neglected the effective mass
of the scalar quantum, as well as the quark masses. With the vector and
axial-vector coupling of $Z^0$ to $u$ and $d$ quarks given
by the standard model, for $\sin^2\theta_W\sim 0.23$, $
A=4\left((g_V^u)^2 + (g_A^u)^2) + (g_V^d)^2 + (g_A^d)^2)\right) \sim \frac{3}{2}$ and
$B= 4 (g_A^u)^2 + (g_A^d)^2 = \frac{5}{4}$.
The Fermi constant is $G_F$. The elementary cross section rises sharply away   
from threshold at $4E^2=m_Z^2$. The last term in the square bracket arises from the
axial-vector coupling of $Z$. This term is responsible for an unusual behavior at
very large energies. The elementary cross section does not fall as $1/E^2$. The
mass $m_Z$ brings in the natural scale, well above which unusual behavior in the
form of an excess becomes predominant. For the purpose of illustrating this
behavior in a transparent manner, we fold the elementary differential cross section 
with a simple parametrization \cite{ref10} of the $\ol{q}-q$ luminosity
function $(\frac{dL}{d\tau})_{u\ol{u}} = 4 (\frac{dL}{d\tau})_{d\ol{d}} = 
\frac{2e^{-10\sqrt{\tau}}}{\tau}$. (There is time to perform folding with complete
parametrizations of quark distributions if the above experimental indications
are confirmed, and if this dynamical idea proves to be useful in relating the effects.)
The folding with eq.~(1) results in an illustrative formula for the transverse momentum
$(p_T)$ distribution for the spin-zero jet quantum produced in $\ol{p}+p\to Z+J_0$,
\ba
&&\frac{d\sigma_{J_0Z}}{dp_T}({\mathrm picobarns/GeV})= 
8\times 10^4 \left( \frac{g_0^2}{4\pi}\right)\frac{p_T}{s}
\int_{\tau_{min}}^1 d\tau \frac{e^{-10\sqrt{\tau}}}{\tau^2}
\left(1-\frac{m_Z^2}{\tau s} \right)^{-1}\nn\\
&\times&\left[1-\frac{4p_T^2}{\tau s \left(1-\frac{m_Z^2}{\tau s}\right)^2}\right]^{-\frac{1}{2}}
\left\{ \frac{3}{32}\frac{m_Z^2}{p_T^2}\left(1-\frac{m_Z^2}{\tau s}\right)^2
-\frac{5}{16}\left(\frac{m_Z^2}{\tau s} - \frac{1}{2}\right)\right\}
\ea
with $\tau_{min} = (p_T + \sqrt{p_T^2 + m_Z^2})^2/s$. The scalar jet quantum $J_0$ could
break up via two gluons; therefore it could exhibit a tendency to materialize into
hadrons as two (correlated) jets. For this reason, we have used for the elementary
cross section the full formula including the squared, effective mass, $m_{J_0}^2$,
of the complete hadronic system which results from the break-up and fragmentation
of the spin-zero quantum $J_0$. This is given by\footnote{The present results correct the cross-section
formulae in ref.~\cite{ref9}}
\ba
\frac{d\sigma_{J_0Z}}{dy} &=&  \sqrt{2}G_{\mathrm F} m_Z^2 \frac{4}{9}\frac{g_0^2}{4\pi}
\frac{\left(x^2-\frac{m_{J_0}^2}{4E^2}\right)^{\frac{1}{2}}}{E^2}\left[ (A+B)\left(\frac{1}{D^2}
+\frac{1}{\ol{D}^2}\right)(1-y^2)\right.\nn\\
&\times&\left(E^4x^2-E^2m_{J_0}^2\right) + 2(A-B)\frac{1}{D\ol{D}}\left(
E^4x^2(1-y^2)+E^2m_{J_0}^2 y^2\right)\nn\\
&+&  2 (A+B)\left(\frac{(u^2-m_{J_0}^2u + \frac{m_{J_0}^4}{4})}{D^2} + \frac{(v^2-m_{J_0}^2 v + 
\frac{m_{J_0}^4}{4})}{\ol{D}^2}\right)\left(\frac{E^2}{m_Z^2}\right)\nn\\
&+& \left. 4(A-B) \frac{(uv - \frac{m_{J_0}^2}{2}(u+v) + \frac{m_{J_0}^4}{4})}{D\ol{D}}\left(\frac{E^2}{m_Z^2}\right)\right]
\ea
with 
\ba
D=(2u-m_{J_0}^2) &\;\;\;\;\;\;& \ol{D} = (2v-m_{J_0}^2)\nn\\
u=E^2x \left(1-\left(x^2-\frac{m_{J_0}^2}{E^2}\right)^{1/2} \left(\frac{y}{x}\right)\right)&&
v=E^2x \left(1+\left(x^2-\frac{m_{J_0}^2}{E^2}\right)^{1/2} \left(\frac{y}{x}\right)\right) \nn\\
x=\left(1+\frac{m_{J_0}^2}{4E^2}-\frac{m_Z^2}{4E^2}\right)&& y=\cos\theta\nn
\ea
In fig.~3, we exhibit the $p_T$-distribution $\frac{d\sigma_{J_0Z}}{dp_T}$ at
$\sqrt{s}=1.8\TeV$ and in fig.~3 the $p_T$-distribution at $\sqrt{s}=630\GeV$, for several values of 
$m_{J_0}^2$. Solely for comparison, we have also shown in figs.~2,3 the $p_T$-distribution 
$\frac{d\sigma_{gZ}}{dp_T}$  for $\ol{p}+p\to Z^0 + g$, calculated approximately from the 
elementary cross section (with $m_g^2=0$) 
\ba
\frac{d\sigma_{gZ}}{dp_T} &=& \sqrt{2}G_{\mathrm F} \frac{4}{9}\alpha_s(p_T) \left(1-\frac{m_Z^2}{4E^2}
\right)^{-1} \frac{(A/2)}{(1-y^2)} \nn\\
&\times& \left(\left(1+\frac{m_Z^2}{2E^2}\right)^2 + \left(1-\frac{m_Z^2}{4E^2}\right)^2 y^2\right)
\ea
In this illustrative numerical work, we use a fixed scalar coupling,\newline 
$\frac{g_0^2}{4\pi} \sim 2(\alpha_s(m_Z))^2 \sim 0.025$; and for $g$, we use the 
decreasing (with scale $p_T$) coupling\newline 
$\alpha_s(p_T^2)\sim\frac{1.64}{\ln(25p_T^2)}(1-\frac{0.66\ln(\ln(25p_T^2))}{\ln(25p_T^2)})$,
($\Lambda=200\MeV$). From the figures, one can follow the onset of a relatively significant
contribution to the differential cross section, at the subpicobarn per GeV level. This
occurs for $p_T$ above $\sim m_Z$. The more flat $p_T$-distribution in production of $J_0$ 
reflects the more flat angular distribution in the CM of the basic $\ol{q}-q$-collision process. 
The enhancement at quite large $p_T>2m_Z$, is substantial. However, the level of the cross
section is then very low, $<10^{-1}\;\mbox{pb/GeV}$. The quark vertex for an effective scalar
quantum may involve two-gluon structure; in any case, damping is to be expected,
at high energies, for emission of $J_0$ with very large momenta, thus reducing this cross section.
In fig.~4, at $\sqrt{s}=1.8\TeV$, we plot the $p_T$-distribution $\frac{d\sigma_{J_0Z}}{dp_T}$
for the case in which $\left(\frac{g_0^2}{4\pi}\right)$ is taken as decreasing like $(\alpha_s(p_T/2))^2$,
instead of being a fixed number $\sim 0.025$, as in fig.~2. Such a ``running'' coupling strength
implies, of course, a non-perturbative domain at low momenta (see footnote F2 below).
From fig.~4 (relative to fig.~2), one notes an increase in the $p_T$ at which
excess jet production begins to stick out prominently, to $p_T\sim 1.5m_Z$,
and a reduction in the cross section.
In fig.~6, we plot an integrated cross section $\sigma_{J_0Z} =
\int_a^b dp_T(\frac{d\sigma_{J_0Z}}{dp_T})$ versus $\sqrt{s}$, where we use $a=50 \GeV$ and
where $b=(p_T)_{max} = \left\{\left(\frac{s+m_{J_0}^2-m_Z^2}{2\sqrt{s}}\right)^2 - m_{J_0}^2\right\}^{1/2}$. 
This cross section is shown for the several values of $m_{J_0}^2$.
At $\sqrt{s}=1.8\TeV$, $\sigma_{J_0Z}\simlt \sigma_{gZ}/2$; thus the integrated effect is not
large. (The detailed expectation from $g$ alone has a comparable uncertainty arising from the choice
of scale \cite{ref5}, and the present data have a similar uncertainty\cite{ref5}.)\par
We are motivated to examine the possible role of this dynamical mechanism in bringing about unusual
behavior in deep-inelastic lepton-nucleon scattering (DIS) by the following considerations.
For neutral-current DIS, the relevant Feynman diagrams are shown in fig.~8. The $(-Q^2)$ must
be sufficiently large in order to make the matrix element from exchange of a virtual $Z^0$ 
comparable to that from exchange of a virtual photon. This brings in the sizable
axial coupling of $Z^0$ to quarks.  Concerning the other two relevant variables \cite{ref2}:
\begin{itemize}
\item[(1)] the scaled energy loss from the lepton, in the rest system of the proton,
$y$ $(0\le y \le 1)$, tends to be preferentially large because of  the explicit momentum factor
in the phase space for $J_0$, which favors backward scattered electrons in the laboratory,
(this is so in a perturbative domain, $\left(\frac{g_0^2}{4\pi}\right)\ll 1$).$^{{\mathrm F}2}$
\item[(2)] the intermediate-state quark lines in fig.~8 involve, in general, virtual quarks,
not mass-shell quarks with $m^2$ taken as zero. For diagram (a), in particular, this means
$(Q+q_i)^2 = (p_{J_0}+q)^2 \ge 0$. Thus, $2Q \cdot q_i \ge (-Q^2)$, which converts into 
$(xy)\ge \frac{(-Q^2)}{s}$ ($\sqrt{s}=300$ GeV at HERA; the present excess of events \cite{ref2}
involve $(-Q^2)\approx (2m_Z)^2$). Since $y_{max}=1$, $x$ must be greater than 0.36. The 
value of $x$ is then determined by the effective invariant mass of the all of the hadrons
that are to be included in the final-state system (jet or jets, plus possible relatively
widely-spread, softer systems\footnote{If a nonperturbative domain occurs for very low
momenta of a scalar quantum, then the condition $xy\to \frac{-Q^2}{s}$ with $(-Q^2)\gg m_Z^2$
as the relevant physical condition for bringing $Z$-exchange and the axial-vector coupling
to the fore, tends to favor the larger values of $x$ for which a significant initial quark
flux is still present; thus also large values of $y$.}).
\end{itemize}
Calculation of the squared matrix element arising from the two diagrams in fig.~8 exhibits
a dominant piece involving only the axial-vector coupling, with the form\footnote{The matrix 
elements from the two diagrams in fig.~8 interfere. There is a prominent piece when
the intermediate-state quark lines have a comparable virtuality, that is, 
$(Q+q_i)^2\sim -(q_i-p_{J_0})^2$.}
\be
2\left( 2(g_A^u)^2 + (g_A^d)^2\right)\sim 1
\ee
The term causes an analogous unusual behavior to that encountered in the process
$\ol{q}+q\to Z+J_0$. It brings in an excess of jet production which can become 
noticeable when values of $(-Q^2)\gg m_Z^2$ are attained, because the axial-vector
coupling comes to the fore. A rough estimate indicates a cross section significantly
less that $0.1\;\mbox{pb}$ at the quark level\footnote{A cross section of 0.1 pb 
may be considered as the smallest value which can explain the most notable of the
anomalous events in the experiments of ref.~\cite{ref2}; that is, the two 
events in each experiment which have $(-Q^2)>30000 (\GeV)^2 \sim (2m_Z)^2$.}, using
the small number $\left(\frac{g_0^2}{4\pi}\right)\sim 0.025$. A more significant
effect requires increasing this number, at least for low momenta$^{\mathrm F2}$ of
a hypothetical scalar quantum\footnote{This may suggest an effective scalar coupling
strength which ``runs'' upward to of order unity in a non-perturbative domain
at low $p_{J_0}$}. Therefore, it is worth emphasizing that there are general
issues connected with the possibility of dynamical jet enhancement.  
 One concerns
the characteristics of the hadronic jet, or jets - that is the kinematic aspects of all
of hadrons included in this system. Are there aspects of these hadrons which differ from
jets arising from single, ordinary quarks? In particular, production of $J_0$ together 
with the quark, in an overlapping jet structure, can give rise to a significant spreading
of particles in the $(x-z)$ plane (or in the $(y-z)$ plane), that is in pseudorapidity;
also a spreading in $(x-y)$ plane, that is in azimuthal angle. The possibility that a 
spin-zero, jet-producing quantum breaks down via two gluons is a motivation to look for
unusual multi-jet structure - three jets in ${\ell}-p$ DIS; four jets in $e^+e^-$ annihilation.
There can be a significant component of heavy quarks associated with $J_0$. Perhaps
the ``bottom'' line of these considerations is that the nature of the jet 
\underline{structure} associated with possible anomalous behavior at high energy 
and momentum transfer, must be investigated in the data.\par
In processes involving ``strong'' interactions alone, $J_0$ production gives rise
to an increased inclusive jet production. The effect need not be dramatic\cite{ref4}.
As an illustration, in fig.~7 ($\sqrt{s} = 1.8\TeV$) and fig.~8 ($\sqrt{s} = 630 \GeV$) 
we examine additional two-jet processes by plotting a ratio of differential cross sections
$R(p_T,\sqrt{s})$ versus $p_T$, where
\be
R(p_T, \sqrt{s}) = \frac{\left(\frac{d\sigma(J_0,g)}{dp_T} + \frac{d\sigma_q(g,g)}{dp_T}\right)}
{\frac{d\sigma_q(g,g)}{dp_T}}
\ee
The process $\ol{q}+q\to J_0+g$ utilizes eq.~(1) with $m_Z^2\to 0$; the process $\ol{q}+q\to g+g$
utilizes eq.~(4) with $m_Z^2\to 0$ (and with appropriate coupling parameters and color factors).
The \underline{axial-vector} couplings are \underline{absent}. The fixed 
coupling $\left(\frac{g_0^2}{4\pi}\right)$ is taken as 0.025, and $\alpha_s(p_T)$ 
decreases with increasing $p_T$. Note that if, at $\sqrt{s}=630\GeV$, one were to
renormalize $R$ to unity at a value of the scaled variable $x_T\sim 0.35$
($E_T\sim 110 \GeV$), then at the same $x_T$, the quantity is above unity at
$\sqrt{s}=1.8\TeV$ ($E_T\sim 315\GeV$). Observe behavior of this kind in
preliminary data in figs.~10 and 11 of ref.~7. The presence of scalar quanta
can make the gluon coupling strength $\alpha_s(p_T)$ appear to ``run'' downward
more slowly, when data is parameterized by $\alpha_s(p_T)$ alone. In fig.~9,
we show $R(p_T,\sqrt{s}=1.8\TeV)$ for the case in which $\left(\frac{g_0^2}{4\pi}\right)$
is taken as decreasing like $(\alpha_s(p_T/2))^2$. Corresponding considerations
can be made for the process of production of direct photons plus
one and two jets. Certain discrepancies exist in preliminary data at the
Tevatron\cite{ref11}; in particular, note section 4 of ref.~11 concerning two jets.
A general consideration is that correlation of two jets originating in a scalar
quantum can reduce the relative occurrence of the larger transverse momenta of a jet,
as measured relative to the effective mass of the full photon-jet-jet system. There
can be an excess of events at low momenta (mass).\par
We conclude by simply noting that the present experimental indications for
``non-standard'' behavior in certain processes at high energy and momentum transfer, which
involve hadronic jets in the final state, may be correlated\footnote{To our
knowledge, the present work is the only paper which addresses the possibility
of a new physical connection between several tentative, unusual effects
at the Tevatron, and the apparently anomalous behavior at HERA. We have become aware
of a recent paper, ref.\cite{ref13}, which discusses some of the Tevatron data
and the HERA data. This paper involves additional precise adjustment (in which the authors are 
authorities) of the parton distributions at large $x$, with an unusual form, 
so that the usual QCD calculations produce ``excess'' events at high energy 
and $p_T$. This is, in a sense, complementary to the present work,
where we have considered new dynamics in the elementary cross sections, and have used a qualitative parton
flux for illustration of the physical effects. (See also ref.~\cite{ref14}, involving leptoquarks.)}. 
The reason for this lies
in the possibility that the jet production itself involves new dynamics, rather than
being a secondary product of the production of many \cite{ref12} exotic particle states \cite{ref2}.\par
One author (S.~B.~) thanks Paolo Giromini for communications from Fermilab. This author also
thanks Patrick Heiliger for help.

\newpage
\section*{Figures}
\begin{figure}[h]
\begin{center}
\mbox{\epsfysize 12cm \epsffile{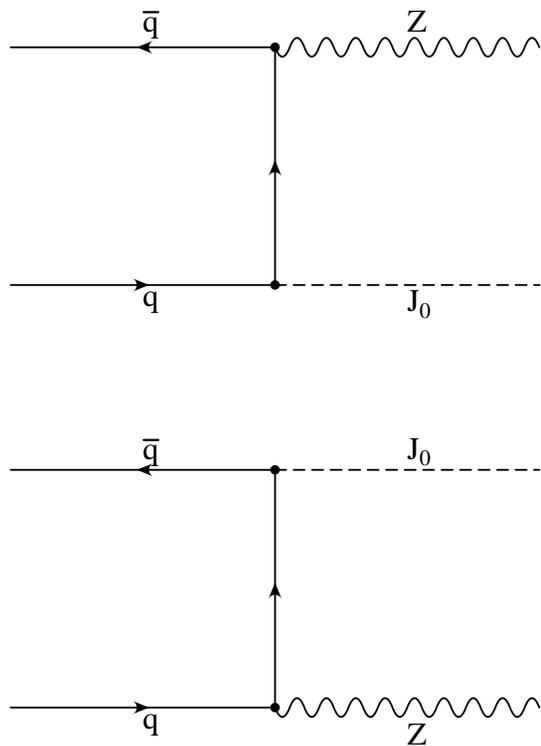}}
\caption{Feynman diagrams for the process $\ol{q}+q\to Z+J_0$. The interference between
the two diagrams is instrumental in giving unusual behavior in the cross section due
to the axial-vector coupling to quarks of $Z$.}
\end{center}
\end{figure}
\begin{figure}[t]
\begin{center}
\mbox{\epsfysize 12cm \epsffile{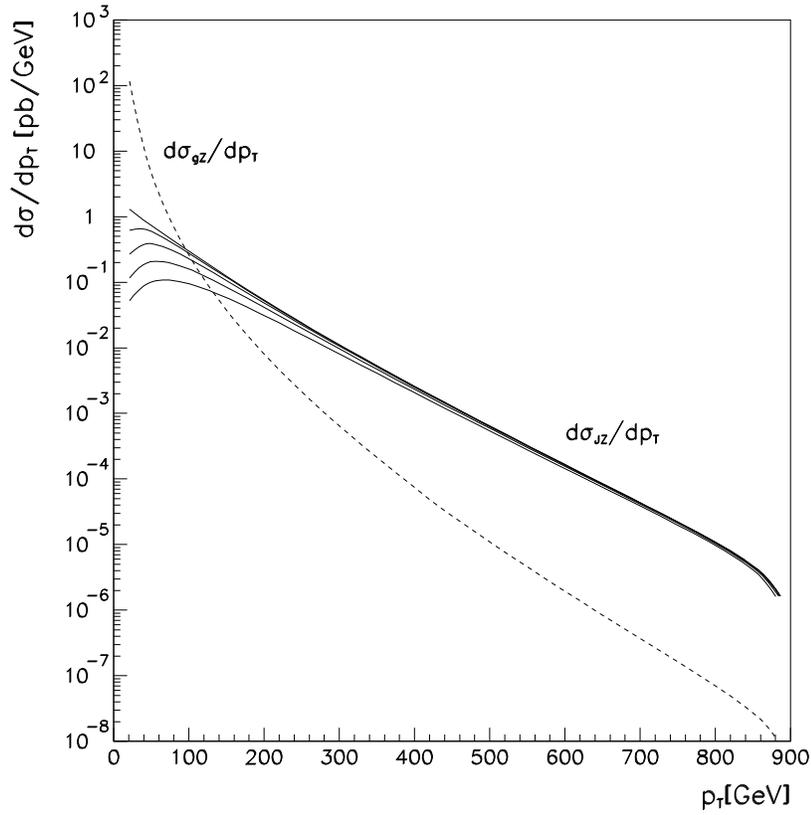}}
\caption{The solid curve gives $\;d\sigma_{J_0Z}/dp_T$ versus $p_T$ for $\ol{p}+p\to Z+J_0$ at 
$\protect\sqrt{s}=1.8\TeV$. $\left(\frac{g_0^2}{4\pi}\right)=0.025$. 
Five values of the effective mass of the hadronic jet from $J_0$
are considered: $m_{J_0} =$ 0, 30, 60, 100 and 150 GeV, reading from the upper curve
to the lower. Solely for comparision, the dashed curve gives $d\sigma_{gZ}/dp_T$ for
$\ol{p}+p\to Z+g$ at $\protect\sqrt{s}=1.8\TeV$, as calculated approximately in the text ($m_g^2 = 0$).}
\end{center}
\end{figure}
\begin{figure}[t]
\begin{center}
\mbox{\epsfysize 12cm \epsffile{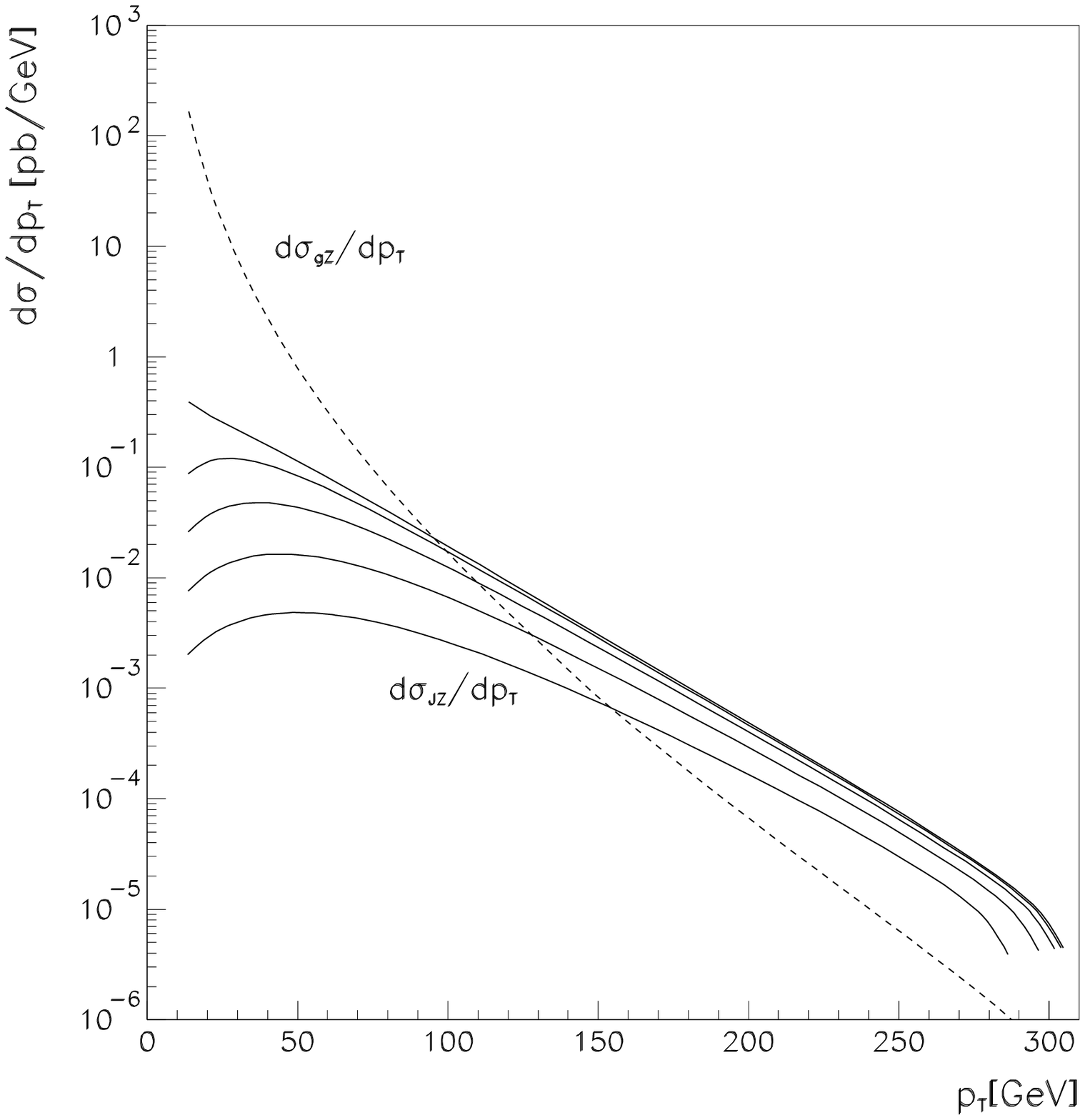}}
\caption{The same as in fig.~2, but at $\protect\sqrt{s}=630\GeV$.}
\end{center}
\end{figure}
\begin{figure}[t]
\begin{center}
\mbox{\epsfysize 12cm \epsffile{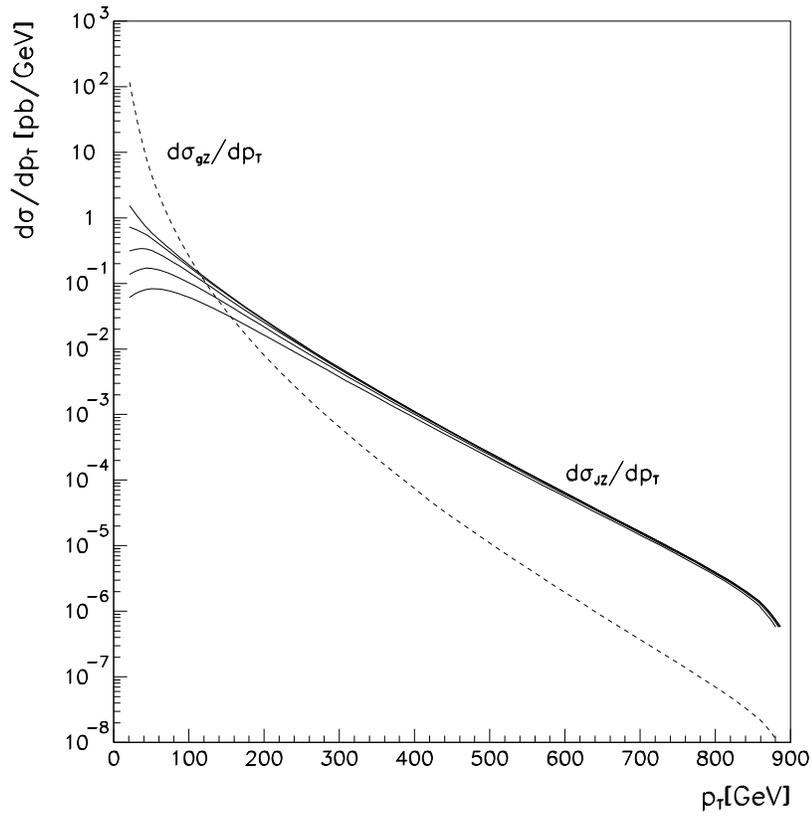}}
\caption{The same as in fig.~2, but with $\left(\frac{g_0^2}{4\pi}\right)$ taken as a function
decreasing like $(\alpha_s(p_T/2))^2$ for increasing scale $p_T$.}
\end{center}
\end{figure}
\begin{figure}[t]
\begin{center}
\mbox{\epsfysize 12cm \epsffile{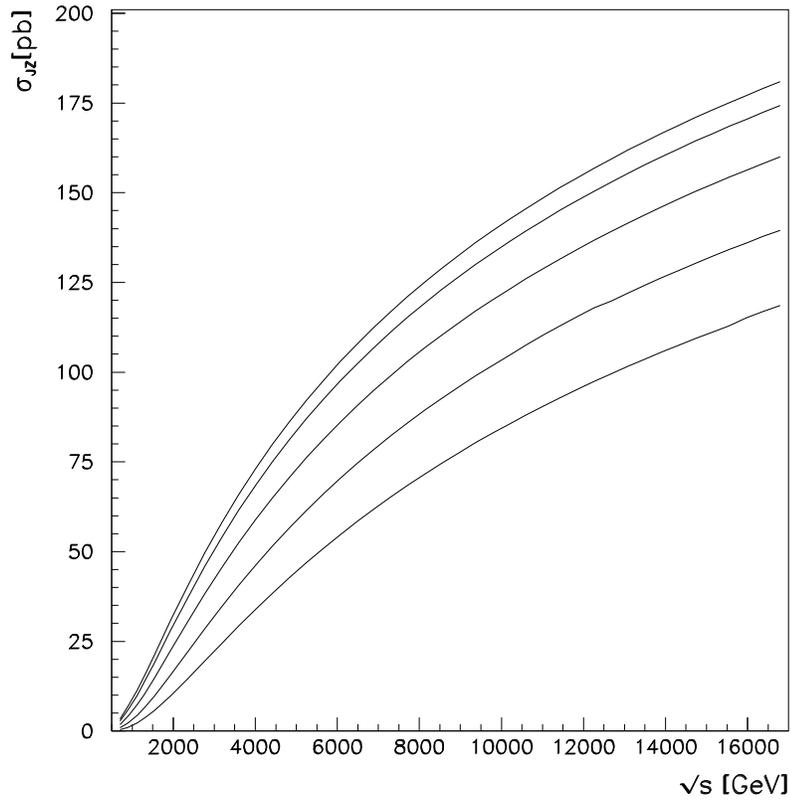}}
\caption{An integrated cross section $\sigma_{J_0Z}(s) = \int_a^b dp_T (\frac{d\sigma_{J_0Z}}{dp_T})$ versus
$\protect\sqrt{s}$, for $a=50\GeV$, 
$b=(p_T)_{max} = \left\{\left(\frac{s+m_{J_0}^2-m_Z^2}{2\protect\sqrt{s}}\right)^2 - m_{J_0}^2\right\}^{1/2}$.
 $\left(\frac{g_0^2}{4\pi}\right)=(\alpha_s(p_T/2))^2$.}
\end{center}
\end{figure}
\begin{figure}[t]
\begin{center}
\mbox{\epsfysize 12cm \epsffile{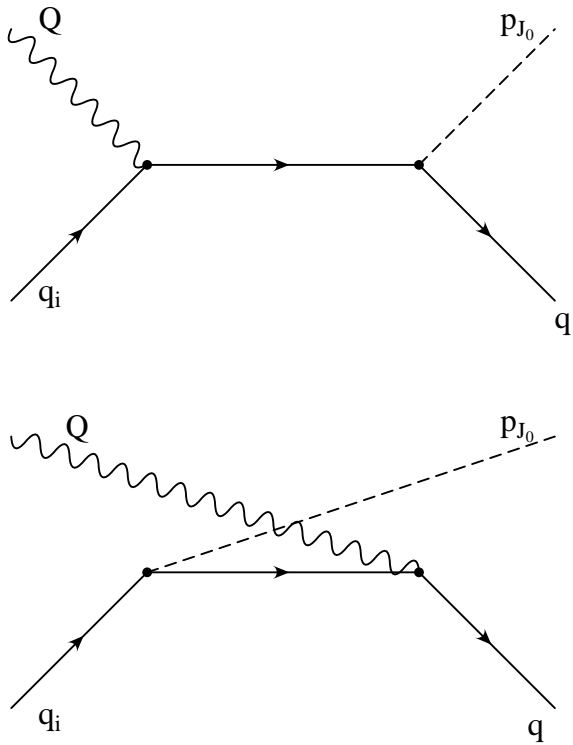}}
\caption{Feynman diagrams relevent to $Z$-exchange in deep-inelastic lepton-nucleon scattering. The
four-momenta of the quanta are indicated. A virtual $Z$ with high $(-Q^2)$ is exchanged to an initial
quark with $q_i$ $(q_i^2\approx 0)$. The final-state contains $J_0$ with $p_{J_0}$, and a quark with 
$q$ $(q^2\approx 0)$.}
\end{center}
\end{figure}
\begin{figure}[t]
\begin{center}
\mbox{\epsfysize 12cm \epsffile{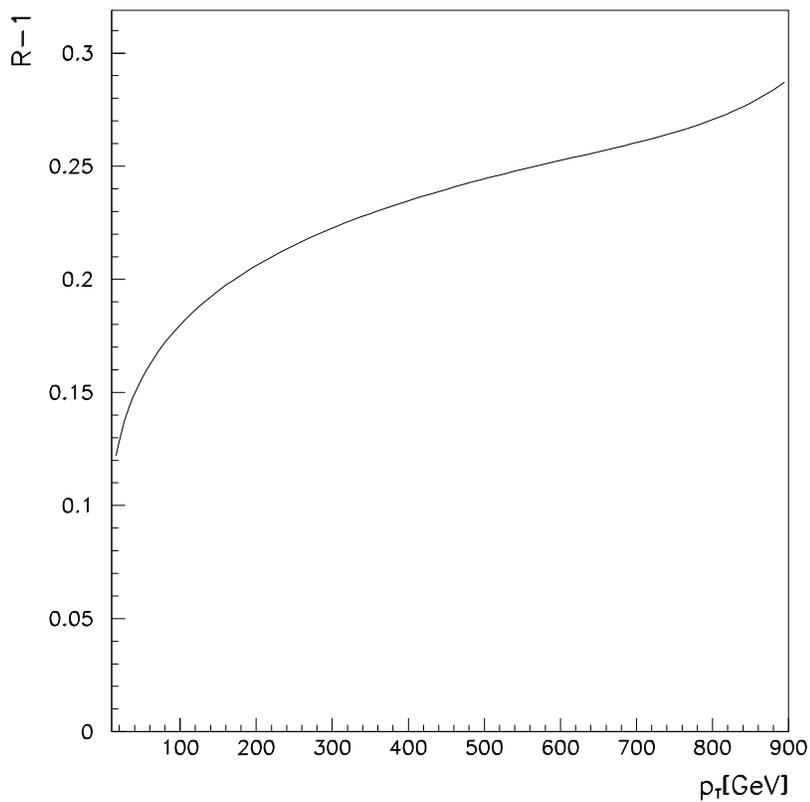}}
\caption{The ratio of differential cross sections, $R(p_T,\protect\sqrt{s})$ given in eq.~(6) of the text,
at $\protect\sqrt{s}=1.8\TeV$, using $\left(\frac{g_0^2}{4\pi}\right)\sim 0.025$. The rise of $R$ above
unity provides an estimate of an excess inclusive jet cross section.}
\end{center}
\end{figure}
\begin{figure}[t]
\begin{center}
\mbox{\epsfysize 12cm \epsffile{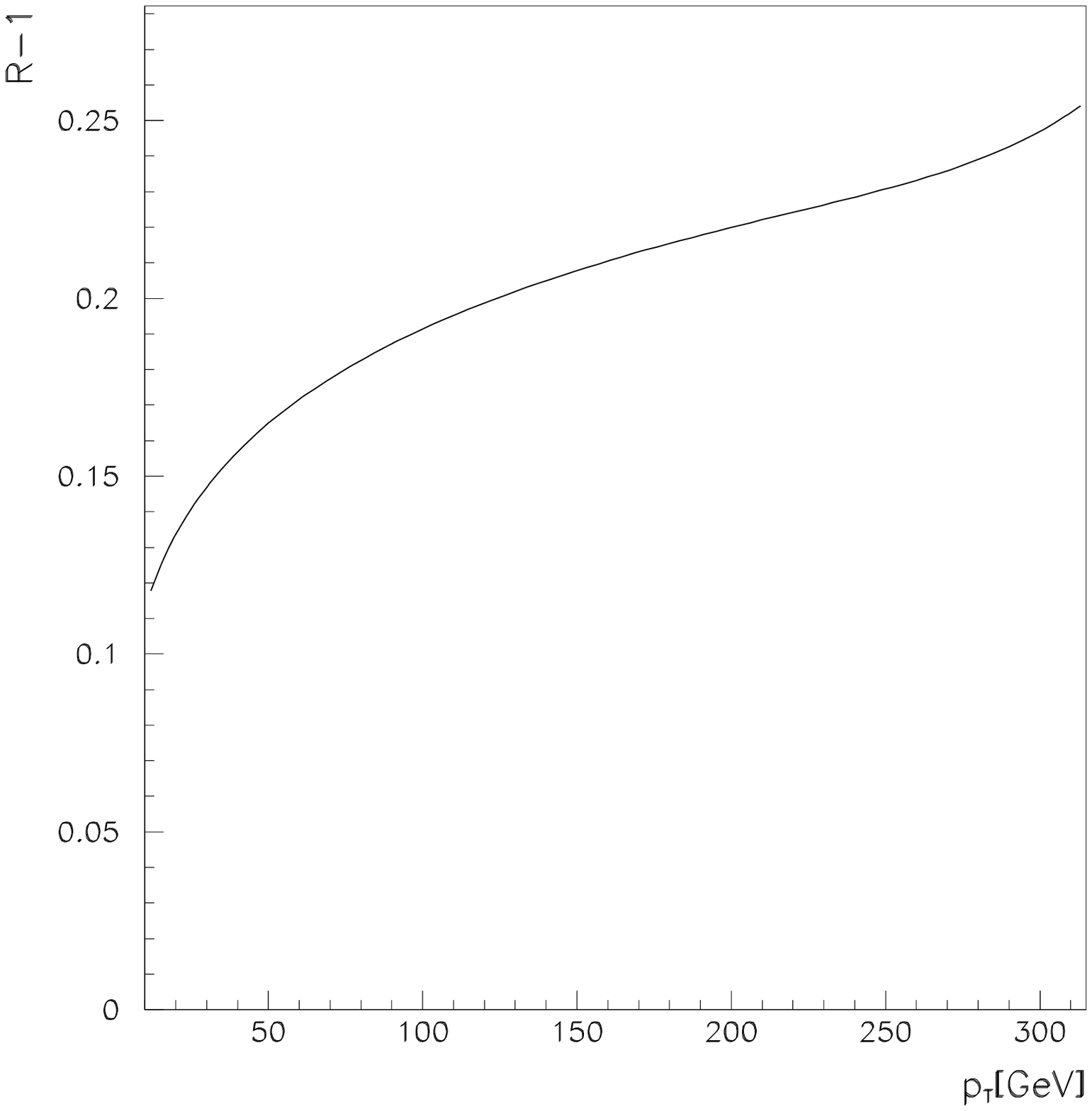}}
\caption{The same as in fig.~9, but at $\protect\sqrt{s}=630\GeV$.}
\end{center}
\end{figure}
\begin{figure}[t]
\begin{center}
\mbox{\epsfysize 12cm \epsffile{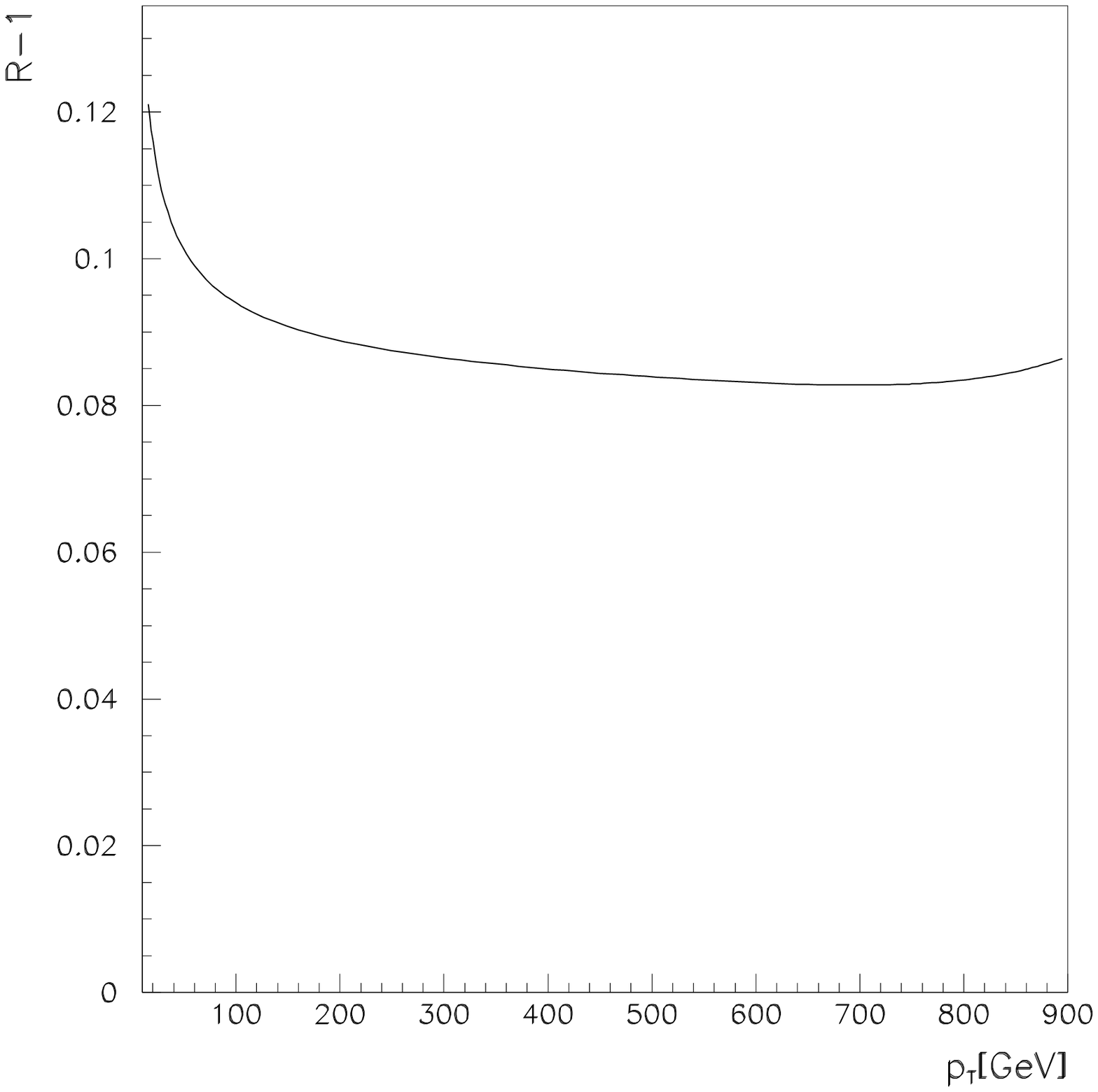}}
\caption{The ratio $R(p_T,\protect\sqrt{s})$ at $\protect\sqrt{s}=1.8\TeV$, 
but with $\left(\frac{g_0^2}{4\pi}\right)$
taken as a function decreasing like $(\alpha_s(p_T/2))^2$ for increasing scale $p_T$.}
\end{center}
\end{figure}

\begin{thebibliography}{99}
\bibitem{ref5} Private communications from Paolo Giromini concerning data
of the CDF Collaboration, which is under analysis.
\end{thebibliography}
\end{document}